\documentclass[aps,superscriptaddress,superbib,twocolumn,groupedaddress]{revtex4}

\usepackage[highlight]{spin-textures}

\bibpunct{}{}{,}{s}{}{}

\def\figurename{Fig.}

\usepackage[paperwidth=8.5in, paperheight=11in]{geometry}
\usepackage{graphics} 
\usepackage{times}
\usepackage{bbm}
\usepackage{tabularx}
\usepackage{tabulary}
\usepackage{tabu}
\usepackage{url}
\usepackage{float}
\usepackage{pdfpages} 

\newcommand{\CoNi}[2]{Co\textsubscript{$#1$}Ni\textsubscript{$#2$}}

\begin{document}

\title{A material view on extrinsic magnetic domain wall pinning in cylindrical CoNi nanowires}

\author{M.~Sch\"{o}bitz}
\email{michael.schobitz@gmail.com}
\affiliation{Univ.\ Grenoble Alpes, CNRS, CEA, Spintec, Grenoble, France}
\affiliation{Friedrich-Alexander Univ.\ Erlangen-N\"{u}rnberg, Chemistry of Thin Film Materials, Erlangen, Germany}
\affiliation{Univ.\ Grenoble Alpes, CNRS, Institut N\'{e}el, Grenoble, France}
\author{O.~Novotn\'y}
\affiliation{Univ.\ Grenoble Alpes, CNRS, CEA, Spintec, Grenoble, France}
\author{B.~Trapp}
\affiliation{Univ.\ Grenoble Alpes, CNRS, Institut N\'{e}el, Grenoble, France}
\author{S.~Bochmann}
\affiliation{Friedrich-Alexander Univ.\ Erlangen-N\"{u}rnberg, Chemistry of Thin Film Materials, Erlangen, Germany}
\author{L.~Cagnon}
\affiliation{Univ.\ Grenoble Alpes, CNRS, Institut N\'{e}el, Grenoble, France}
\author{C.~Thirion}
\affiliation{Univ.\ Grenoble Alpes, CNRS, Institut N\'{e}el, Grenoble, France}
\author{A.~Masseb\oe uf}
\affiliation{Univ.\ Grenoble Alpes, CNRS, CEA, Spintec, Grenoble, France}
\author{E.~Mossang}
\affiliation{Univ.\ Grenoble Alpes, CNRS, Institut N\'{e}el, Grenoble, France}
\author{O.~Fruchart}
\affiliation{Univ.\ Grenoble Alpes, CNRS, CEA, Spintec, Grenoble, France}
\author{J.~Bachmann}
\email{julien.bachmann@fau.de}
\affiliation{Friedrich-Alexander Univ.\ Erlangen-N\"{u}rnberg, Chemistry of Thin Film Materials, Erlangen, Germany}
\affiliation{Institute of Chemistry, Saint-Petersburg State Univ., St.\ Petersburg, Russia}

\date{\today}

\begin{abstract}
Speed and reliability of magnetic domain wall (DW) motion are key parameters that must be controlled to realize the full potential of DW-based magnetic devices for logic and memory applications. A major hindrance to this is extrinsic DW pinning at specific sites related to shape and material defects, which may be present even if the sample synthesis is well controlled. Understanding the origin of DW pinning and reducing it is especially desirable in electrochemically-deposited cylindrical magnetic nanowires (NWs), for which measurements of the  fascinating physics predicted by theoretical computation have been inhibited by significant pinning.
We experimentally investigate DW pinning in Co$_x$Ni$_{100-x}$ NWs, by applying quasistatic magnetic fields. Wire compositions were varied with $x=20,30,40$, while the microstructure was changed by annealing or varying the pH of the electrolyte for deposition. We conclude that pinning due to grain boundaries is the dominant mechanism, decreasing inversely with both the spontaneous magnetization and grain size. Second-order effects include inhomogeneities in lattice strain and the residual magnetocrystalline anisotropy. Surface roughness, dislocations and impurities are not expected to play a significant role in DW pinning in these wire samples.
\end{abstract}

\maketitle


\section{Introduction}

The motion of domain walls~(DWs) has been a point of high interest in magnetism since the early 20\textsuperscript{th} century\cite{bib-SIX1931}, as it is often involved in magnetization reversal processes. Extensive efforts have been made to comprehend the underlying physics and its link with sample shape and microstructure. One-dimensional nanosized conduits such as strips and wires are a textbook case to model and therefore understand DW motion, owing to the small number of degrees of freedom involved\cite{bib-BEA2005,bib-THI2006,bib-MOU2007,bib-BOU2011}. The rise of spintronics brought prospects for a wide range of applications for DW motion in such conduits. Some prime examples include the dynamic switching of magnetic tunnel junctions or spin valves~\cite{bib-GRI2020}, the ultra fast speeds achieved in compensated ferrimagnets\cite{bib-CAR2018} or the progress made towards realizing a non-volatile shift-register memory device\cite{bib-PAR2004,bib-PAR2008,bib-FER2017}. 
Across all of these, reliability and wall velocity stand out as critical parameters governing viability. While the physics of DW motion is now rather well understood, both reliability and velocity may suffer from imperfections in material and shape, called extrinsic pinning sites\cite{bib-KLA2005,bib-BOG2009,bib-IVA2011,bib-HER1990,bib-FRU2019b}. These act as energy wells or barriers that a DW must overcome\cite{bib-KON1937,bib-KON1940} to move, translating into a minimum driving force that must be applied to allow for wall motion past each pinning site, such as a critical depinning field, $\Hdep$, or critical depinning current, $\jdep$, for the case of DW motion induced by the application of an external magnetic field or spin-transfer torque, respectively.


The detailed nature of extrinsic pinning sites is unclear even in simple one-dimensional magnetic systems. Various ideas have been put forth, including compositional changes along the sample length\cite{bib-VOG2011,bib-IVA2016b,bib-MOH2016}, polycrystallinity and associated grain boundaries\cite{bib-HER1990,bib-LIN2015}, surface defects and roughness\cite{bib-KLA2005,bib-IVA2011}, dislocations\cite{bib-YU1999}, impurities\cite{bib-FER2013d} or strain\cite{bib-DEA2011,bib-LEI2013}, which all lead to local variations in magnetic properties such as magnetocrystalline anisotropy, spontaneous magnetization, exchange stiffness, \etc{} and therefore variations in the energy landscape felt by the DW. 
%


Among one-dimensional magnetic conduits, there exists a specific interest in cylindrical ferromagnetic nanowires (NWs), synthesized by electrochemical deposition into porous membranes\cite{bib-BOC2017}. Indeed, theory and simulations predict fascinating novel physics in these 3D systems\cite{bib-FRU2017b}, such as DW velocities in excess of \SI{1}{km/s} with no Walker breakdown\cite{bib-THI2006,bib-WIE2010}, only limited by the emission of spin waves (spin-Cherenkov effect)\cite{bib-YAN2011b,bib-HER2016}. We recently provided a first experimental evidence of the very high velocities\cite{bib-FRU2019b} in CoNi NWs, however, the results were affected by DW pinning, resulting in a large spread in the results. The pinning field in cylindrical NWs is typically one order of magnitude higher than in thin flat strips of the same material however deposited by physical methods\cite{bib-HAY2006b}. This obstacle must be overcome in order to further study the fascinating fast motion of DWs and associated effects in these systems.

A vast quantity of reports concern the coercivity of NWs\cite{bib-SEL2001,bib-NIE2002,bib-HER2004,bib-QIN2005, bib-PIT2011, bib-VIV2012, bib-IVA2013b, bib-PER2013, bib-FRU2018b}. In most cases, especially for rather soft-magnetic materials, this relates to the nucleation of DWs from a wire end, shown to depend only on the geometry and materials parameters. For instance, coercivity in \CoNi{x}{100-x} NWs shows a minimum of nucleation field for composition $x=40$\cite{bib-PER2013,bib-BOC2017}. However, the nucleation field in such NWs is larger than $\Hdep$, so that the DW runs the entirety of the wire once nucleated, and nothing is learned about extrinsic pinning through coercivity.
Far fewer studies have focused on pinning. These largely theoretical and simulation-based works have considered DW pinning on diameter\cite{bib-DOL2014,bib-BER2016,bib-NAS2019,bib-FRU2018c,bib-FRU2020} or composition modulations\cite{bib-IVA2016b,bib-MOH2016,bib-BRA2018}, on surface roughness\cite{bib-IVA2011, bib-DOL2014} and on regions with different magnetocrystalline anisotropy, which can be likened to a polycrystalline texture\cite{bib-IVA2011,bib-FRA2011}. All successfully identify that these material defects cause DW pinning, and some relate them primarily to variations in magnetostatic and anisotropy energy.  However, only Ivanov and Orlov\cite{bib-IVA2011} provide a detailed theoretical picture of DW pinning due to surface roughness and polycrystallinity. A limited number of experimental pieces of work have focussed on pinning sites fabricated on purpose, through composition\cite{bib-MOH2016} or diameter modulations\cite{bib-BER2016,bib-NAS2019}. Other possible explanations for DW pinning, such as grain boundary pinning, have not been supported experimentally.






In this paper we address the issue of DW pinning in cylindrical NWs with rather soft-magnetic properties, by evaluating experimentally the impact of various types of defects on DW pinning in \CoNi{x}{100-x} NWs. We consider compositions in the range of $20$ to \SI{40}{\%} Co with a view to exhibit a low-pinning material, since we may expect low pinning from \CoNi{40}{60} NWs that exhibit the lowest coercive field\cite{bib-BOC2017}, or from \CoNi{20}{80} material that exhibits near-zero magnetocrystalline anisotropy and magnetostriction\cite{bib-KAD1981}.


\section{Methods}

For the electrochemical deposition of NWs, templates of porous anodized aluminium oxide (AAO) membranes were synthesized as described elsewhere\cite{bib-BOC2017} and a coating of Au was sputtered on a single side to provide an electric contact for electrodeposition. We fabricated samples with compositions \CoNi{20}{80}, \CoNi{30}{70} and \CoNi{40}{60} starting from modified Watt's electrolytes with compositions summarized in Table~\ref{tab:electrolytes}. The electrolyte pH was set to $2.5$ for all depositions, except once lowered to $1.5$ to reduce the NW grain size with respect to the standard sample\cite{bib-NAT1996,bib-ALP2008}, by adding NaOH or H\textsubscript{2}SO\textsubscript{4}. Depositions were carried out in a three-electrode electrochemical cell at \SI{-1.1}{V} \textit{vs.} an Ag/AgCl/NaCl (\SI{3}{M}) reference electrode. After NW growth, the gold contact was etched in a KI/I\textsubscript{2} solution. Annealing of some batched was carried out to allow for grain growth and crystallographic relaxations\cite{bib-QIN2002,bib-TOT2010,bib-WAN2004}, by placing filled membranes into a tubular vacuum oven at \SI{500}{\degree C} for \SI{20}{min}.

\begin{table*}[t]
	\begin{center}
		\begin{tabularx}{\textwidth}{*{7}{>{\centering\arraybackslash}X}}
		& \multicolumn{2}{c}{\CoNi{20}{80}} & \multicolumn{2}{c}{\CoNi{30}{70}} & \multicolumn{2}{c}{\CoNi{40}{60}}\\
		Compound & $n$ (mol) & $m$ (g) & $n$ (mol) & $m$ (g)  & $n$ (mol) & $m$ (g)\\
		\hline
		CoSO\ensuremath{_4\cdot}7H\ensuremath{_2}O& 0.0182 & 0.5104 & 0.0282 & 0.7930 & 0.0450 & 1.2649\\
		CoCl\ensuremath{_2\cdot}6H\ensuremath{_2}O & 0.0012 & 0.0276 & 0.0018 & 0.0428 & 0.0027 & 0.0649\\
		NiSO\ensuremath{_4\cdot}6H\ensuremath{_2}O&  0.3120 & 8.2005 &  0.2934 & 7.7109 & 0.2375  & 6.2426\\
		NiCl\ensuremath{_2\cdot}6H\ensuremath{_2}O &  0.0551 & 1.3087 & 0.0518 & 1.2306 & 0.0400 & 0.9508\\
		H\ensuremath{_3}BO\ensuremath{_3} & 0.5000 & 3.0915& 0.5000 & 3.0915& 0.5000 & 3.0915 \\
		saccharin & 0.0150 & 0.2750  & 0.0150 & 0.2750 & 0.0150 & 0.2750\\
		\end{tabularx}
		\caption{Composition of various electrolytes and mass per \SI{100}{mL} used to electrochemically grow NWs with a given composition. Low pH depositions used the same electrolyte composition, resulting in slower growth rates but similar NW compositions.} \label{tab:electrolytes}
	\end{center}
\end{table*}

NW compositions were checked by atomic absorption spectroscopy (AAS) and by scanning electron microscopy energy-dispersive x-ray (SEM EDX) analysis. Vibrating sample magnetometry (VSM) and x-ray diffraction (XRD) were carried out on large ($\SI{2x2}{mm^2}$) pieces of filled AAO membrane. For XRD, diffractograms were obtained with a symmetric $\theta-2\theta$ geometry and a Copper filament as x-ray source, with characteristic $\mathrm{K}_\alpha$ radiation with a wavelength of $SI{1.5402}{\AA}$. Individual diffraction peaks in a diffractogram were fitted using a Gaussian to extract both peak position and peak width.

Unless stated otherwise, NWs were freed by dissolving the AAO membrane in \SI{0.6}{M} chromic and \SI{0.4}{M} orthophosphoric acid solution at \SI{70}{\degree C} for $3$~hours. The NW-acid solutions were rinsed several times in distilled water and finally in ethanol to produce clean NW suspensions. Drops of suspension were then dispersed onto silicon wafers with prepared gold alignment marks and NWs were located using SEM. These marks allowed to seek the wires were in the optical view of the magnetic force microscope (MFM), which was used to monitor DW motion. Individual NW samples were also imaged with transmission electron microscopy~(TEM), by dispersing NWs on a specific lacey-carbon grid.


\section{Results}

\begin{figure}
\centering\includegraphics{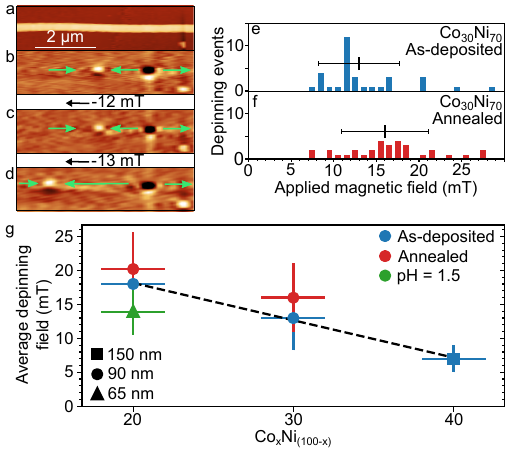}
\caption{DW depinning in NWs. (a) AFM image of a section of an as-deposited \SI{90}{nm}-diameter \CoNi{30}{70} NW, with the initial magnetic configuration of the same section shown in the MFM image in (b). There are two DWs present and the magnetization of the longitudinal domains is indicated by green arrows. (c, d) MFM images taken after the application of $-12$ and \SI{-13}{mT} fields in the indicated direction, respectively, showing the depinning and subsequent pinning of the left hand DW at \SI{-13}{mT}. (e, f) Distributions of depinning fields for \CoNi{30}{70} as-deposited and annealed NWs, respectively, with the average value and standard deviation indicated by the horizontal lines. (g) Average depinning field as a function of NW cobalt content, for as-deposited NWs (blue), annealed NWs (red) and as-deposited NWs deposited with an electrolyte pH~$= 1.5$ (green). NW diameters vary from \SI{150}{nm} (squares) to \SI{90}{nm} (circles) to \SI{65}{nm} (triangles) and the dashed line acts as a guide to the eye.}
\label{fig:MFM}
\end{figure}

DW depinning was measured by imaging DW positions with MFM before and after applying an external magnetic field oriented along the NW long axis, with pulse duration of about~\SI{1}{\second}. By slowly increasing the amplitude of the applied field between each measurement, the field required to depin DWs from specific pinning sites could be determined. \subfigref{fig:MFM}{a} shows an atomic force microscopy (AFM) image of a section of an individual, as-deposited \SI{90}{nm}-diameter \CoNi{30}{70} NW. The corresponding MFM image in \subfigref{fig:MFM}{b} shows the initial magnetic configuration with two DWs located along the wire length, and the direction of magnetization within the three longitudinal domains indicated by green arrows. The MFM image in \subfigref{fig:MFM}{c} shows the magnetic configuration after applying a magnetic field pulse with amplitude $\SI{-12}{mT}$, remaining unchanged. Imaging after applying a $\SI{-13}{mT}$ field pulse\bracketsubfigref{fig:MFM}{d} shows that the left-hand wall was depinned and moved to a different, presumably stronger pinning site, hence the depinning field of the initial site is in the range \SIrange{12}{13}{mT}.  Such measurement series were repeated on multiple NWs for each sample type in order to determine a distribution of depinning field amplitudes.

\subfigref{fig:MFM}{e,f} shows distribution of depinning field for both as-deposited and annealed \CoNi{30}{70} NWs, respectively. Note that the procedure of applying quasistatic pulses of field with rising values during a series implies that for any given wire contributing to the statistics in these figures, once the first depinning event has occured we cannot measure events associated with lower pinning fields. Thus, these distributions probably over-emphasize pinning sites associated with a high value of field, partly contributing to their large width. Still, the average and standard deviation of the distributions are plotted as horizontal lines, illustrating a lower average depinning field of the as-deposited sample. The variation of average depinning field as a function of cobalt content and different preparation methods is presented in \subfigref{fig:MFM}{g}, with different NW diameters indicated by marker shapes.
In the case of the as-deposited NW samples (blue), the average depinning field decreases from $18\pm{4}$ to $13\pm{4}$ to $7 \pm \SI{2}{mT}$ as the Co content increases from $20$ to $30$ to \SI{40}{\%}. This nearly linear decrease is indicated by the dashed guide to the eye. NWs deposited from an electrolyte with a pH of $1.5$\bracketsubfigref{fig:MFM}{g, green} exhibit a lower average depinning field compared to the NWs deposited from a normal electrolyte with a pH of $2.5$ (blue). Conversely, NWs that were annealed in the AAO membrane at \SI{500}{\degree C} for \SI{20}{min}\bracketsubfigref{fig:MFM}{g, red} display a larger average depinning field (data from \subfigref{fig:MFM}{f}).
%

\begin{figure}
\centering
\includegraphics{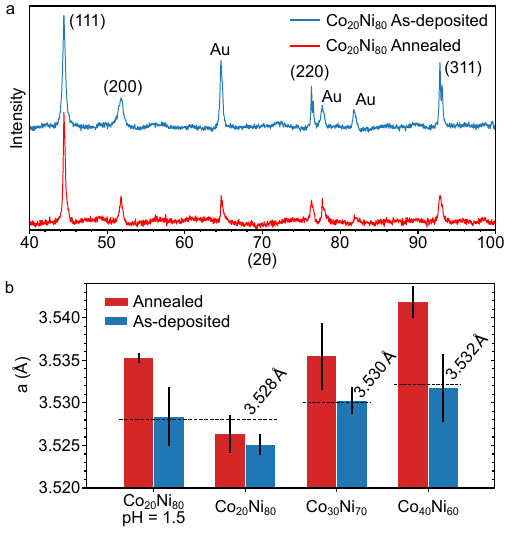}
\caption{XRD analysis of NW crystallography. a) Individually rescaled XRD diffractograms of \CoNi{20}{80} NWs enclosed in an AAO membrane pre (blue) and post (red) annealing at \SI{500}{\degree C} for \SI{20}{min}. The peaks corresponding to face-centered cubic CoNi are indexed, while the remaining peaks at $64.6$, $77.7$ and \SI{82.6}{\degree} result from the remainder of the gold bottom electrode. b) Lattice coefficients, $a$, in \AA{}, calculated from XRD peak positions for all samples, pre (blue) and post (red) annealing. Theoretical lattice parameters calculated with Vegard's law for the given composition are indicated by the dashed lines.}
\label{fig:XRD}
\end{figure}

It is evident from \subfigref{fig:MFM}{g} that NW samples prepared by different methods lead to variations in average depinning fields. To tentatively separate the effects of composition and microstructure between the samples, we used XRD to analyze the various NWs while in the AAO membranes.
The XRD diffractogram of as-deposited \CoNi{20}{80} NWs is shown in blue in \subfigref{fig:XRD}{a}. The peaks at $2\theta = 44.51$, $51.78$, $76.35$, \SI{92.92}{\degree} are consistent with reflections from the (111), (200), (220), and (311) planes of face-centered cubic CoNi alloy, while the peaks at $64.6$, $77.7$ and \SI{82.6}{\degree} are consistent with diffraction from gold, suggesting that the etching of the bottom electrode is incomplete. The absence of any other peak indicates the single phase of the electrodeposited NWs.

After subtraction of the calibrated instrumental offset, the diffraction peak angles were used to calculate the lattice coefficient, $a$, of the sample's cubic unit cell, according to Bragg's law, giving an average $a=3.525\pm\SI{0.001}{\AA}$\bracketsubfigref{fig:XRD}{b}. This is slightly below the expected value of the lattice coefficient for \CoNi{20}{80}, $a_{\mathrm{th}} = \SI{3.5279}{\AA}$, calculated using Vegard's law, indicating a general compression of the unit cell with respect to the equilibrium state, at least along the wire axis. 
Annealing of the \CoNi{20}{80} NWs sample does not change the XRD pattern\bracketsubfigref{fig:XRD}{a, red} qualitatively, however, a similar calculation of the lattice coefficient gives $a=3.526\pm\SI{0.002}{\AA}$ and thus possibly an increase in unit cell size although affected by the error bar.
At this stage we need to discuss the impact of the AAO membrane. Heating of the AAO and its concomitant thermal expansion leads to a reduction of the pore diameter. This and the NW's own thermal expansion cause a radial stress of the wire which would tend to cause an expansion along the wire axis. This is reversed upon cooling, however, the original shape may not be entirely recovered, leaving a difference with the non-annealed sample. Further, as the sample is quickly removed from the oven, the rapid cooling that occurs could be linked to quenching and partly freeze the expanded high-temperature structure\cite{bib-YU1999}.
\subfigref{fig:XRD}{b} shows the lattice coefficients calculated from XRD peak positions of all samples, with the dashed lines indicating $a_{\mathrm{th}}$ for the given composition\cite{bib-TOT2010}. For both the as-deposited (blue) and the annealed samples (red), $a$ increases with increasing Co content, as a result of the added volume of the larger Co atoms, in agreement with the increasing $a_{\mathrm{th}}$.
All as-deposited samples, except the previously discussed \CoNi{20}{80}, have a near perfect match between $a$ and $a_\mathrm{th}$, and it thus follows that the crystallites within the electrodeposited material are essentially strain-free. 
Furthermore, annealing leads to an increase in $a$, however, unlike the previously discussed case of annealed \CoNi{20}{80}, the other three instances of annealing (\CoNi{20}{80} with pH~$= 1.5$, \CoNi{30}{70} and \CoNi{40}{60}) lead to $a > a_{\mathrm{th}}$, indicative of crystal lattices under slight tensile stress along the NW axis.
Finally, \CoNi{20}{80} NWs electrodeposited from an electrolyte with a pH of $1.5$ exhibit a larger lattice coefficient that is closer to $a_{\mathrm{th}}$ than standard growth \CoNi{20}{80} NWs. This may be a result of the significantly slower deposition rate, allowing atoms to rearrange and grow overall less strained crystals.

\begin{figure}[h!]
\centering
\includegraphics{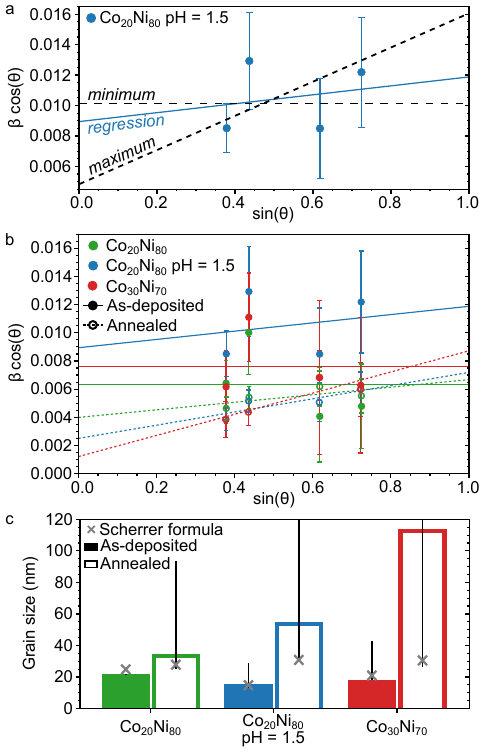}
\caption{Williamson-Hall analysis of different NW samples, with angles expressed here in radiant. a) Williamson-Hall plot for as-deposited \CoNi{20}{80} with pH~$= 1.5$ NWs. The linear regression fit (blue line) and maximum and minimum possible slopes which provide the error (black dashed lines), are indicated. b) The same Williamson-Hall plot for as-deposited (full circles) and annealed (open circles) \CoNi{20}{80} (green), \CoNi{20}{80} with pH~$= 1.5$ (blue) and \CoNi{30}{70} (red) NWs. Linear regression fits are shown as solid and dashed lines for as-deposited and annealed samples, respectively, the y-intercept of which are inversely proportional to the grain size of the sample. c) Grain size (nm) as calculated by the Williamson-Hall method for as-deposited (solid bar) or annealed (open bar) NW samples. Solutions from the Scherrer formula are indicated by grey crosses. Note that the error bars of the annealed \CoNi{20}{80} with pH~$= 1.5$ and annealed \CoNi{30}{70} extend to over \SI{1}{\micro m} and therefore much past the limit of the instrumental precision.}
\label{fig:WH}
\end{figure}

Further microstructural information can be obtained using the Williamson-Hall approach\cite{bib-WIL1953,bib-MOT2012}, to separate the effects of distribution of inhomogeneous crystal strain, $\epsilon$, and grain size, $D$, on peak broadening, $\beta$, of the XRD diffractogram. This method is more suitable than relying on the more basic Scherrer formula, which can underestimate the grain size, affected by $\epsilon$ and instrumental resolution. Indeed, both $\epsilon$ and $D$ contribute to the finite width of peaks, the first by a spread of each diffraction peak and depending on the angle, $2\theta$, the second through a fixed peak width increase scaling as $1/D$ (Scherrer formula). In other words, the effect of $\epsilon$ increases at large $2\theta$, whereas the effect of $D$ remains constant.
Therefore, after correcting for instrumental peak broadening in the XRD diffractogram, the impact of $\epsilon$ and $D$ can be separately identified by following $\beta \cos{\theta}=4\epsilon\sin{\theta}+{k \lambda }/{D}$, with $\lambda$ the x-ray wavelength and $k$ a shape factor best approximated as $0.9$.
A Williamson-Hall plot for as-deposited \CoNi{20}{80} with pH~$= 1.5$ NWs is shown in \subfigref{fig:WH}{a}. From the above equation, the slope ($4\epsilon$) and y-intercept (${k \lambda }/{D}$) of the blue linear regression of the data set provide the amplitude of strain distribution and grain size, respectively. The error for this information is given by the maximum and minimum possible slopes, as shown by dashed lines, however, with the minimum slope not below zero since $\epsilon < 0$ is unphysical.
The full Williamson-Hall plot for both as-deposited and annealed NWs is given in \subfigref{fig:WH}{b}, showing the linear regression fits of as-deposited and annealed samples as solid and dashed lines, respectively. For the as-deposited \CoNi{20}{80} and \CoNi{30}{70} NW data the regression slope should be negative from the numerical fit, however bounded to the zero slope as constraint for its physical meaning, as mentioned above.

\subfigref{fig:WH}{c} shows the grain size for both as-deposited (full bar) and annealed (open bar) NWs calculated from the Williamson-Hall plot. The error bars are calculated by the same method using the maximum and minimum possible fitted lines, however, note that the error bars for annealed \CoNi{20}{80} with pH~$= 1.5$ and \CoNi{30}{70} extend to $> \SI{1}{\micro m}$, which exceeds the instrumental limit for peak widths below \SI{0.02}{\degree} in $2\theta$ and thus $D_\mathrm{max} \lesssim \SI{400}{nm}$. Grain sizes calculated using the Scherrer formula, $D = { k \lambda }/{ \left( \beta \cos{\theta} \right)  }$, again with the shape factor $k=0.9$, should provide a lower bound for grain size, as discussed previously. These are indicated by grey crosses, which match indeed well with the lower bound of the error bars.
The grain size of standard as-deposited NWs is $\approx \SI{20}{nm}$, with little change in composition. However, it must be noted that the large data spread and uncertainties in the Williamson-Hall plot may affect the accuracy of these calculated values.
Decreasing the electrolyte pH reduces the grain size to \SI{15}{nm} (\CoNi{20}{80} with pH~$= 1.5$), possibly resulting from the increased hydrogen evolution at the cathode. This inhibits grain growth by limiting surface diffusion of adatoms and instead favours grain nucleation\cite{bib-RIE2009b}. Conversly, annealing increases the grain size to \SI{35}, \SI{55} and \SI{114}{nm} for \CoNi{20}{80}, \CoNi{20}{80} with pH~$= 1.5$ and \CoNi{30}{70}, respectively, due to the recrystallization and growth initiated by the heat treatment\cite{bib-QIN2002,bib-TOT2010,bib-WAN2004}.

Similarly, the distribution of inhomogeneous crystal strain was calculated from the slope as $4\epsilon$. Where the linear regression slope was positive, the strain distribution was calculated as 0.00067 (annealed \CoNi{20}{80}), 0.00074 (\CoNi{20}{80} with pH~$= 1.5$), 0.00117 (annealed \CoNi{20}{80} with pH~$= 1.5$) and 0.00188 (annealed \CoNi{30}{70}), however, for all measurements the error bar extends to nearly one order of magnitude larger values. Considering the large error associated with these results, it is not possible to draw definite conclusions about the strain distribution in these NW samples.

The grain sizes and strain distributions that we measure are very similar to values for nanocrystalline nickel thin film, also electrodeposited from a Watt's or purely sulphate electrolyte. A large number of studies made use of the Scherrer formula and found grain sizes of the order of $10$ to \SI{50}{nm}, for deposition conditions close to the ones used in this work~\cite{bib-RAS2008,bib-RAS2009,bib-EBR1999,bib-KAN2009,bib-MIS2004}. Fewer studies have investigated electrodeposited NiCo alloys~\cite{bib-TOT2010,bib-PAR2005}, but the reported grain sizes largely match those of pure Ni. In particular, Tóth \etal~\cite{bib-TOT2010} investigated the effect of annealing at \SI{300}{\degree C} for \SI{1}{h} on the grain size of nanocrystalline CoNi alloys and found an increase from $10$ to \SI{40}{nm} for low Co contents. The experiment that is most similar to the present study was performed by Dost \etal~\cite{bib-DOS2020}, where annealing of \SI{275}{nm}-diameter Ni nanowires at \SI{650}{\degree C} for \SI{1}{h} increased the grain size from $8$ to \SI{160}{nm}. TEM imaging also revealed that after annealing grains often occupied the entire diameter of the wire.
The grain sizes of our as-deposited materials match well with values from literature, however, a direct comparison of the impact of annealing in this study and previous literature cannot be made, because the starting material and annealing recipes differ. Regarding the grain size of deposits made with a lower electrolyte pH, there are no reports on nanocrystalline CoNi and other studies investigate pH~$>2$. Still, in the CoFe alloys studied by Riemer \etal~\cite{bib-RIE2009b}, a $50$ to \SI{30}{nm} grain size reduction was reported when changing the pH from $3$ to $2$ and in the NiCu alloys studied by Alper \etal~\cite{bib-ALP2008} a $120$ to \SI{90}{nm} grain size reduction was reported when changing the pH from $3.3$ to $2$.
Strain distribution is considered less often than grain size, especially since most studies discuss only the Scherrer formula to analyse XRD peak broadening. The studies that do report on strain distribution in nanocrystalline Ni films find $\epsilon\lesssim0.005$~\cite{bib-WAN2004,bib-MIS2004} and Wang \etal~\cite{bib-WAN2004} note that this reduces by \SI{30}{\%} by annealing at \SI{100}{\degree C} for \SI{1}{h}.

\begin{figure}[h!]
\centering
\includegraphics{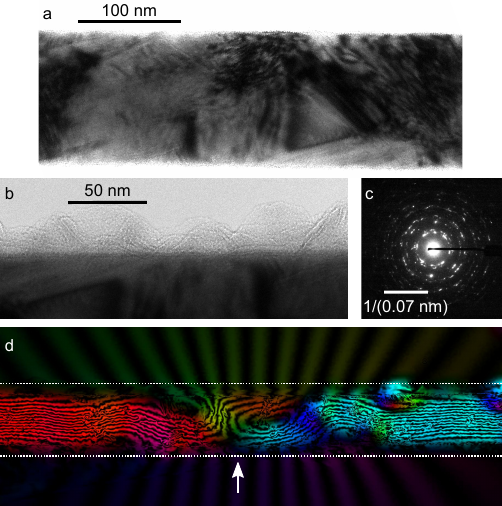}
\caption{TEM images of a \SI{130}{nm}-diameter \CoNi{40}{60} nanowire freed from the AAO membrane by dissolution in NaOH solution to correlate domain wall pinning sites, surface roughness and microstructure. a,b) High resolution TEM images, showing in (a) the complex polycrystalline microstructure and in (b) a higher magnification on the edge of the same nanowire, revealing low surface roughness but strong oxide layer formation from dissolution in NaOH. c) A diffraction pattern from the same type of \CoNi{40}{60} nanowire. d) Electron holography image showing the magnetic flux reconstruction in- and outside the nanowire (outlined by white dotted lines), displayed as black lines superimposed on color map obtained from the electron phase gradient (color associated to in-plane direction). A domain wall (white arrow) is present at the same place observed with conventional imaging in (a).}
\label{fig:TEM}
\end{figure}

\section{Discussion of the origin of pinning}
DW pinning should not occur in a homogeneous singlecrystalline NW with a perfect cylindrical shape. However, such a wire is unachievable in practice, and any deviation from perfection should lead to DW pinning. This includes diameter modulation and surface roughness, polycrystallinity and the associated grain boundaries, strain, changes in composition, line and point defects such as dislocations and impurities. Based on the results presented in the previous section, we evaluate below the phenomena most likely responsible for DW pinning, grouping them in three categories: i) surface roughness and other shape defects; ii) polycrystallinity and its interplay with magnetocrystalline anisotropy and inverse magnetostriction; iii)~material defects such as inhomogeneities, grain boundaries, dislocations, impurities. General aspects applying to those different situations are described first, in the paragraph below.

\subsection*{General considerations}

To set orders of magnitude, the models discussed are applied to a wall of length $\wallWidth\simeq 2R$, a reasonable scaling law for the wire diameters considered here, in the range a few tens to a hundred of nanometers\cite{bib-FRU2015b}. We will also use the textbook model of Becker-Kondorski\cite{bib-KON1937,bib-BEC1939} for domain-wall pinning in a one-dimensional framework, relating the depinning field $\Hdep$ to the energy landscape~$\mathcal{E}(x)$:
\begin{equation}
\Hdep=\frac{1}{2\muZero\Ms S}\fracdiff{\mathcal{E}}{x},
\label{eq-BeckerKondorski}
\end{equation}
with $S$ the cross-sectional area of the one-dimensional conduit, $\pi R^2$ in the present situation, and $\mathcal{E}$ the position-dependent net energy of the DW, labeled in Joules.

\subsection*{Surface roughness}
Here we evaluate to which extent deviations from perfect translational symmetry in the shape of a nanowire may explain the pinning of DWs in our systems. Deviation from the perfect cylindrical shape may take the form of local modulations of diameter~(\ie, correlated around the wire diameter), or general roughness. Analytical modeling cannot cover the general case of pinning on such defects, and approximations must be made. To provide some generality and therefore robustness in the discussion and comparison with experiments, we consider three models of a very different kind.

A first model may be adapted from Bruno \etal, evaluating the contribution of dipolar energy to magnetic anisotropy in thin films, related to the existence of roughness\cite{bib-BRU1988}. This model predicts a cost for planar anisotropy amounting to a net energy, written
\begin{equation}
\mathcal{E}_\mathrm{Bruno}=\mathcal{S}\,\frac12\muZero\Ms^2\,\frac{\sigma}{4}\left[{1-f\left({2\pi\frac\sigma\xi}\right)}\right].
\label{eq-roughnessBruno}
\end{equation}
$\mathcal{S}$ is the surface area of the thin film considered, $\sigma$ the average deviation for the roughness, $\xi$ the correlation length. $f(0)=1$ with an infinite negative slope, sharply decreasing to $f(0.1)\simeq0.6$ and $f(1)<0.1$. Assuming $(\sigma,\xi)\ll R$, one may neglect long-range correlations on magnetostatic energy, and therefore model the surface of a cylindrical wire with the rolled surface of a thin film. We combine \eqnref{eq-BeckerKondorski} to \eqnref{eq-roughnessBruno} with $\mathcal{S}=2\pi R\delta x$, considering a distance $\delta  x=\xi$ equal to the roughness correlation length, and averaging azimuthally over a number of correlation areas $2\pi R/\xi$, we obtain a formula for the depinning field:
\begin{equation}
H_\mathrm{Bruno}=\Ms\,\frac{\sigma}{8R}\,\sqrt{\frac{\xi}{2\pi R}}\left[{1-f\left({2\pi\frac\sigma\xi}\right)}\right].
\label{eq-roughnessBruno2}
\end{equation}

A second model, proposed by Ivanov and Orlov, considers specifically the cylindrical geometry, with an azimuthally correlated roughness\cite{bib-IVA2011}. The models predicts that
\begin{equation}
\Hdep = 1.4\,\Ms \left({V_\mathrm{d}}/{\delta_\mathrm{w}^3}\right) \sqrt{    \left({\delta_\mathrm{w}}/{2R} \right)  \ln{\left({L}/{\delta_\mathrm{w}}\right)}}.
\label{eq-roughnessIvanov2}
\end{equation}
Here we disregard the square root as $\delta_\mathrm{w}\simeq2R$, and $\ln{\left({L}/{\delta_\mathrm{w}}\right)}$ pertains to the statistical distribution of strength of pinning sites on a long length scale~$L$, an aspect not considered in the other two models. $V_\mathrm{d}=2\pi R\,\sigma\,\delta_\mathrm{w}$ is the volume of defects at the scale of a DW, and $\delta_\mathrm{w}^3=8R^3$. We finally have:
\begin{equation}
H_\mathrm{Ivanov}=1.4\,\Ms\,\frac{\pi}{2}\,\frac{\sigma}{R}\;.
\label{eq-roughnessIvanov2}
\end{equation}

Yet a third model, developed by de~Riz, intends to describe gentle and controlled variations of diameter. The propagation field is close to $2\Ms(\linediff{R}{x})$. However, this model is valid for modulations larger than the wall width, which is opposite to the situation of roughness, so that we cannot simply consider that $\linediff{R}{x}\approx\sigma/\xi$. Instead, as a rule of thumb we may renormalize the impact of the modulation of diameter with the length of the full DW, \ie, with an extra coefficient $\xi/(2R)$. This leads to:
\begin{equation}
H_\mathrm{de\;Riz}=\Ms \frac{\sigma}{R}.
\label{eq-roughnessDeRiz}
\end{equation}

Interestingly, although all three models have a very different fundamental basis and a priori range of application, they all lead to pinning scaling with~$\Ms\,\sigma/R$, simply with a different coefficient, and statistical azimuthal averaging for Bruno's model. This expectation is opposite to our experimental observation that the pinning field decreases with increasing~$\Ms$\bracketfigref{fig:MFM}. This suggests that roughness coupled with dipolar effect is not the leading mechanism of pinning in our case. This is consistent with the fact that wires electrochemically deposited in AAO templates are known to be very smooth, resulting from the amorphous structure of the aluminium oxide, unless modulations of diameter may result from instabilities during the anodisation step~\cite{bib-LEE2010c}. TEM imaging of nanowires similar to those studied here revealed that beyond a possible native oxide layer, no noticeable surface roughness could be observed\bracketsubfigref{fig:TEM}{a,b}. Further, imaging with TEM holography\bracketsubfigref{fig:TEM}{d} showed that a domain wall was pinned along the low roughness wire segment shown already in \ref{fig:TEM}a. Such low impact from pinning due to surface roughness may not be the case for polycarbonate templates, which display a roughness intrinsically linked with the molecular size of the underlying polymers~\cite{bib-COR2010b,bib-BIZ2013}.

\subsection*{Polycrystallinity}

Polycrystallinity leads to spatial variations of magnetocrystalline or magnetoelastic anisotropy energy density~$K$, due to different grain orientations. This converts into a position-dependent energy of a DW, which implies pinning. In magnetically-soft bulk-like systems, this situation can be described with the Herzer model\cite{bib-HER1990,bib-HER1992}, averaging the anisotropy energy over the large number of grains inside a DW. The width of the latter is found self-consistently to scale with~$K^4D^6/A^3$. This model is not suitable for cylindrical nanowires, for which $\delta_\mathrm{W}$ is largely determined by magnetostatics and possibly exchange, scaling approximately with $\approx2R$ in our situation. Instead, we may consider the change of energy $\delta\mathcal{E}$ of a DW upon motion with distance $\delta x=D$, and apply the Becker-Kondorski model\bracketeqnref{eq-BeckerKondorski}. $\delta\mathcal{E}=K\sqrt{N}D^3$, with $N=2\pi R^2/D^2$ the change of number of grains in the DW upon motion. This leads to:
\begin{equation}
H_\mathrm{poly}=\frac{1}{\sqrt{2\pi}}\,\frac{1}{\muZero\Ms}\,\frac{D}{R}\,K.
\label{eq-averageAnisotropy}
\end{equation}
Let us apply \eqnref{eq-averageAnisotropy} to our material with largest magnetocrystalline and magnetoelastic anisotropy, \CoNi{40}{60}. At room temperature, the cubic coefficient for magnetocrystalline anisotropy is $K_1 \simeq \SI{7}{kJ/m^3}$\cite{bib-KAD1975}. The magnetoelastic coupling coefficients $B_1$ and $B_2$ are in the range of $\SI{10e7}{J/m^3}$\cite{bib-KAD1981}. Considering strain $\epsilon\lesssim\num{3e-4}$\bracketfigref{fig:XRD} in the as-grown material, the resulting anisotropy density is $\Kmel\simeq\SI{3}{kJ/m^3}$. \eqnref{eq-averageAnisotropy} applied with $D=\SI{20}{\nano\meter}$, $R=\SI{50}{\nano\meter}$, $\muZero\Ms=\SI{1.17}{\tesla}$ and the above values for density of anisotropy energy leads to pinning fields $\si{\milli\tesla}$ or below. This is an order of magnitude lower than experimental values. Besides, those two sources of anisotropy are expected to vanish almost simultaneously for a composition around \CoNi{20}{80}, while in our wires the pinning field is the highest. Thus, experimental observations are inconsistent with polycrystallinity combined with magnetocrystalline or magnetoelastic anisotropy energy as the source of DW pinning in our nanowires.

%
\subsection*{Material defects}
A potential imperfection in the material are inhomogeneities in composition of the CoNi alloy, leading to an axial gradient of magnetic energy for a~DW. While this may be done on purpose\cite{bib-MOH2016} to engineer DW positions, undesired inhomogeneities may also arise from potential instabilities or limitations of diffusion during electroplating. However, SEM EDX line scans evidenced homogeneous compositions along the wire lengths in all of our studied samples.

A usual defect in materials is dislocations. Yu \etal\cite{bib-YU2000} noticed that coercive fields increase for larger dislocation densities. Lindquist \etal\cite{bib-LIN2015} directly observed pinning on dislocations in bulk magnetite, in which higher dislocation densities and longer dislocation lengths increase depinning fields, while pinning is decreased for larger magnetization. This hints at an inhomogeneity of magnetoelastic anisotropy due to strain around the dislocation, balanced by the pressure induced on the DW by the Zeeman energy. The lowering of the micromagnetic energy around a dislocation has been modeled, promoting nucleation for magnetization reversal\cite{bib-ABR1962}. The increase of $\Ms$ and the experimental decrease of pinning for larger Co content are consistent, however again we would expect very low pinning for the \CoNi{20}{80} alloy with vanishing magnetostriction. Besides, it is known that annealing reduces dislocation lengths and densities\cite{bib-QIN2002,bib-TOT2010,bib-WAN2004} and thus pinning, while we observe an increase of pinning upon annealing\bracketsubfigref{fig:MFM}{e-f}. Therefore, dislocations are probably not the most active source of pinning in our case.

Another kind of material defect is grain boundaries. These are by nature two-dimensional, potentially having a larger impact than dislocation on pinning DW, which are two-dimensional as well. The pinning of DW on grain boundaries could indeed be observed directly, \eg with Lorentz TEM in bulk FeCo alloys\cite{bib-YU1999,bib-YU2000}. The reason for pinning at a grain boundary is the various disruptions in the material, such as mismatch of crystalline lattices, strain and possibly accumulation of impurities, including non-magnetic. Accordingly, micromagnetic material parameters may be inhomogeneous, including magnetization, anisotropy and exchange stiffness. A one-dimensional model of DW shall reasonably well describe such situations. This has been the focus of many reports for the search of explanations of coercivity in practical materials, and tackling the Brown paradox\cite{bib-BRO1963b}. Aharoni considered a step\cite{bib-AHA1960} or a linear decrease\cite{bib-ABR1960} in magnetic anisotropy. This can be generalized to a step\cite{bib-HAG1970} or the local variation\cite{bib-LEI1997} of any micromagnetic parameter. In all cases the propagation field scales inversely with magnetization, as the result of the balance of the Zeeman pressure against a magnetization-independent energy profile. This is a general feature in the physics of coercivity, valid also in other situations such as elastic domain walls bending between pinning sites\cite{bib-DOeR1938,bib-KER1948}. The only exception is pinning originating from dipolar anisotropy on roughness, as examined in the first case above, whose energy depends on magnetization and scale with~$\Ms^2$.

Coming back to our measurements, pinning on grain boundaries is consistent with the decrease in depinning field [$18\pm{4}$ to $13\pm{4}$ to $7\pm\SI{2}{mT}$] with increasing Co content and increasing $\Ms$ in the as-deposited NW samples\bracketfigref{fig:MFM}, which all share similar grain sizes\bracketfigref{fig:WH}. Besides, this explanation is compatible with a pinning persisting for the \CoNi{20}{80} alloy despite its vanishing microscopic sources of anisotropy, unlike the hypothesis of the effect of polycrystallinity examined previously. Still, if the grain boundary area is smaller than the section of the wire, the pinning effect is probably averaged out over multiple grain boundaries. Again, this fact is consistent with the increase of pinning strength upon annealing\bracketfigref{fig:WH}. Not only grain boundaries are expected to become more extended, but annealing also promotes impurity diffusion to the grain boundaries\cite{bib-WAN2002}. The reduced depinning field for NWs deposited with $\mathrm{pH}= 1.5$ [$14\pm\SI{3}{mT}$] matches with the reduced grain size compared to the as-deposited NWs [$18\pm\SI{4}{mT}$]. Similarly, one may expects pinning to decrease with increasing NW diameter, related to the averaging of energy over the number of grains boundaries across the diameter.

\subsection*{Overview and discussion}
To conclude the above analysis of various physical reasons for pinning, the most plausible explanation lies in the role of grain boundaries, able to affect micromagnetic material parameters locally. Surface roughness and random anisotropy in a polycrystalline material probably play a minor role in the present case. Still, the granular structure of the material has an impact, partially averaging the effect for smaller grain size, and thereby decreasing the pinning effect. It is possible to reduce the depinning field by increasing $\Ms$, here through a larger Co content, however, the appearance of the hexagonal close packed Co phase above a composition of \CoNi{50}{50}\cite{bib-VIV2012} may increase pinning due to the increasing magnetocrystalline anisotropy. However, this trend may not apply for current-driven DW motion, for which a higher transfer of angular momentum is required for large magnetization. Initial experiments comparing as-deposited and annealed \SI{90}{nm} diameter \CoNi{30}{70} NWs revealed an increase in depinning current $\jdep$ from $\SI{1.2e12}{A/m^2}$ in as-deposited wires to $\SI{2.3e12}{A/m^2}$ in annealed wires, which matches the trend observed under quasistatic fields. However, it appears that the changes in pinning strength have a much greater impact on $\jdep$ than on $\Hdep$. Finally, although reducing the grain size appears to be an effective method to reducing pinning strength, it also leads to increased resistance in NWs. Therefore, temperature-assisted depinning shall crucially be taken account in the analysis of experiments, and would also imply a higher power consumption in applications.

\section{Conclusion}

We have experimentally investigated DW pinning at room temperature in electrodeposited and rather magnetically-soft \CoNi{x}{100-x} NWs, with $x=20$, $30$ and $40$. Pinning is highest for $x=20$ and decreases linearly with increasing Co content, scaling roughly with $1/\Ms$. This observation is not compatible with pinning governed by surface roughness, which should rather increase in proportion to $\Ms$. Also, this does not support the role of magnetocrystalline anisotropy and magnetostriction, as at room temperature these are lowest nearly simultaneously for the \CoNi{20}{80} concentration. This suggests that pinning is governed by microstructural defects such as grain boundaries, supported by information on grain size and boundaries brought by XRD and TEM, and the increase of pinning for larger grain size~$D$. This suggests a handle to controlling DW pinning in electrodeposited wire by decreasing $D$ through engineering of deposition parameters or conversely increasing $D$ by annealing.

\section{Acknowledgements}
M. S. acknowledges a grant from the Laboratoire d’excellence LANEF in Grenoble (ANR-10-LABX-51-01). The authors acknowledge financial support from the CNRS-CEA “METSA” French network (FR CNRS 3507).


\bibliographystyle{apsrev4-1}
%

\end{document}